\documentclass[amsfonts,twocolumn]{revtex4}
\usepackage{amsmath}
\usepackage[dvips]{epsfig}
\begin{document}

\author{Shmuel Marcovitch and Benni Reznik}
\title{Testing Bell inequalities with weak measurements}
\affiliation{School of Physics and Astronomy,
Raymond and Beverly Sackler Faculty of Exact Sciences,
Tel-Aviv University, Tel-Aviv 69978, Israel.}


\begin{abstract} Quantum theory is inconsistent
with any local hidden variable model as was first shown by Bell.
To test Bell inequalities
two separated observers extract correlations from a common ensemble of identical systems.
Since quantum theory does not allow simultaneous measurements of noncommuting observables,
on each system every party measures a single randomly chosen
observable out of a given set.
Here we suggest a different approach for testing Bell inequalities that is experimentally realizable by current methods.
We show that Bell inequalities can be maximally violated even when
all observables are measured on each member of the ensemble.
This is possible by using weak measurements that produce
small disturbance, at the expense of accuracy.
However, our approach does not constitute an independent test of quantum nonlocality since
the local hidden variables may correlate the noise of the measurement instruments.
Nevertheless, by adding a randomly chosen precise measurement at the end of every cycle of weak measurements,
the parties can verify that the hidden variables were not interfering with the noise, and thus validate the suggested test.
\end{abstract}
\maketitle
\section {Introduction}
Quantum nonlocality \cite{bell} has been tested \cite{aspect,zeilinger} in different systems
and is now widely accepted.
Its most simple manifestation is given by Clauser-Horne-Shimony-Holt (CHSH) inequality \cite{chsh},
in which every cycle $A$ and $B$ measure one out of two randomly chosen observables
$a_1,a_2\in\{\pm1\}$ and $b_1,b_2\in\{\pm1\}$.
In a local hidden variable model $B\leq2$, where
\begin{equation}
\label{chsh}
B=|E(a_1b_1)+E(a_1b_2)+E(a_2b_1)-E(a_2b_2)|.
\end{equation}
Above $E(a_ib_j)$ is the joint expectation value.
The EPR state
\begin{equation}
\label{state}
|\psi\rangle=\frac{1}{\sqrt{2}}(|\uparrow^A_z\uparrow^B_z\rangle+|\downarrow^A_z\downarrow^B_z\rangle),
\end{equation}
saturates the maximal bound $2\sqrt{2}$ \cite{tsirelson} with {\it i.e.}
$a_1=\sigma_x$, $a_2=\sigma_z$, $b_1=\sigma_{\pi/4}$, and $b_2=\sigma_{3\pi/4}$,
where
$\sigma_{\pi/4}=\frac{1}{\sqrt{2}}(\sigma_x+\sigma_z)$ and
$\sigma_{3\pi/4}=\frac{1}{\sqrt{2}}(\sigma_x-\sigma_z)$.

\begin{figure}[ht]
\center{
\includegraphics[width=3.2in]{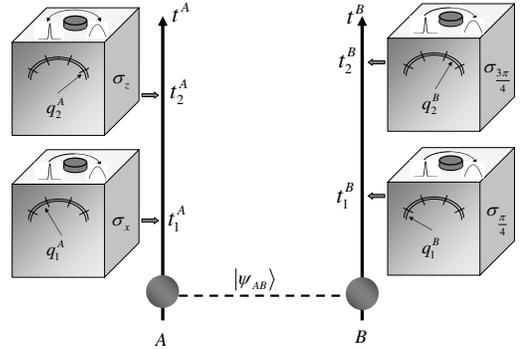}
\caption
{
Sequential CHSH setup. Parties $A$ and $B$ measure two observables
at sequential times.
The knobs control
the inaccuracy of the devices. In the accurate measurement
limit the pointer readings correspond to the possible values of the observables ($\pm1$).
In the inaccurate (weak) limit the readings are noisy, and the
microscopic values cannot be inferred.
We require that the distance between the parties $d=|X_A-X_B|$ satisfies
$d\ll c t_2^A$ and $d\ll c t_2^B$.
}
\label{fig1}
}
\end{figure}

In the regular setting every cycle the parties measure a single observable.
In our approach to be referred to as the {\it sequential setting},
for each member of the ensemble
$A$ measures {\em both} $a_1$ and
$a_2$ (one after another), and similarly $B$ measures both $b_1$ and
$b_2$, as is schematically described in Fig. 1.
Obviously, as discussed above, for ordinary measurements the first measurement of
$a_1$ would randomize the result
in the subsequent measurement of $a_2$,
and no violation of Bell inequalities would be manifested.
However, as is well known, there is a trade-off between
the accuracy of the measurement and the disturbance
caused to the system \cite{von}.
The limit in which individual measurements provide vanishing information gain
was first analyzed by Aharonov {\it et. al.} (in the context of post-selection)
\cite{weak} and was termed {\it weak measurements}.
In this limit
the measurements become inaccurate and one could expect that nonlocal correlations
would not be observed.

The main result of this paper is that as one gradually reduces the disturbance caused to the system,
by decreasing the accuracy of measurements,
one in fact regains the maximal violation of Bell inequalities.
This result is stated in theorem 1 and corollary 1.
The outcomes of inaccurate measurements may lie outside the usual range of the measured observables.
In this case local hidden variables may interfere with the noise to violate Bell inequalities.
Therefore our approach does not constitute an independent test of quantum nonlocality.
However, by adding a single randomly chosen precise measurement at the end of each cycle,
the parties can determine whether the hidden variables interfere with the noise.
The expected negative result implies that the inaccurate measurements still provide a valid test.
This is manifested in corollary 2.
We then proceed by quantitatively illustrating the transition from strong to weak measurements in the sequential setting
in realistic finite ensembles
and show that {\em simultaneously} different Bell tests can be attained
using the same ensemble.
We conclude by discussing implications of our approach,
in particular maximal violation of the Leggett and Garg inequalities \cite{leggett}
and possible realizations of the suggested method.

\section{Violation of Bell inequalities with weak measurements}
{\bf Theorem 1}. {\it Let a system $|\psi\rangle$ be measured sequentially by an arbitrary number of observables $O_i$ ($i=1,\dots,m$)
at times $t_i$ where $0<t_1<t_2<\cdots<t_m$.
In the limit of weak measurements,
\begin{eqnarray}
\label{seq1} &&E_{QM}(q_i)=\langle \psi|O_i(t_i)|\psi\rangle,\\
\label{seq2} &&E_{QM}(q_i q_j)=\mbox{Re}\langle \psi|O_i(t_i)O_j(t_j)|\psi\rangle,
\end{eqnarray}
where
$q_i$ is the reading of the $i$th measurement apparatus pointer.
The observables $O_i$ are written in the Heisenberg representation,
$O_i(t_i)\equiv U^{\dagger}(t_i)O_i U(t_i)$,
$U(t)=e^{-i H_0 t}$, and
$H_0$ is the free Hamiltonian.
}
\\
\\
{\bf Proof.}
The full
Hamiltonian is \begin{equation}
\label{h}
H=H_0+\sum_i^m\delta(t-t_i)p_i O_i,
\end{equation}
where 
$[q_i,p_i]=i$ (we take $\hbar=1$).
The second term in Eq. (\ref{h}) corresponds to the von-Neumann measurement interaction,
where for simplicity instantaneous measurements are assumed.
In addition, we assume identical initial Gaussian wavepackets for the pointers:
\begin{equation}
\phi(q_i,t=0)=\left(\frac{\epsilon}{2\pi}\right)^{1/4}e^{-\epsilon q_i^2/4},
\end{equation}
where $\epsilon^{-1}=[\Delta q_i(t=0)]^2\equiv\sigma^2$.
The initial state of the system and the apparatuses
\begin{equation}
|\Psi\rangle=|\psi\rangle\otimes\prod_i\phi(q_i),
\end{equation}
evolves in time according to
\begin{equation}
{\cal U}=\prod_i{\cal U}_i\equiv\prod_i U^{\dagger}(t_i)e^{-i p_i O_i} U(t_i).
\end{equation}
Each operation of $p$ yields an order of $\epsilon$.
In the limit of weak measurements, the inaccuracy $\sigma\to\infty$ ($\epsilon\to0$).

By expanding ${\cal U}_i$ to first order in $\epsilon$,
while keeping terms to first order in ${\cal U}$ and
tracing out the system and $q_j$, $j\neq i$,
one obtains Eq. (\ref{seq1}).

To derive Eq. (\ref{seq2}) we expand to second order
\begin{equation}
\nonumber
e^{-i p_iO_i}=1-i p_iO_i-\frac{1}{2}p_i^2O_i^2+o(\epsilon^3).
\end{equation}
After the measurements the state of the systems and the pointers is given by
\begin{equation}
\label{expand1}
\begin{split}
|\Psi\rangle\!&=U(t_m)\!\Big[\prod_l^m\!\phi(q_l)\!-\sum_k^m\!\phi'(q_k)\prod_{l\neq k} \phi(q_l)O_k(t_k)\!
\\&+\sum_{k<l}\!\phi'(q_k)\phi'(q_l)\!\prod_{s\neq k,l}\!\phi(q_s) O_l (t_l) O_k(t_k)
\\&+\frac{1}{2}\sum_k^m\!\phi''(q_k)\prod_{l\neq k} \!\phi(q_l)O_k^2(t_k)\!+\!o(\epsilon^3)\!\Big]\! |\psi\rangle.
\end{split}
\end{equation}
Let us compute $E_{QM}(q_iq_j)$, where without loss of generality $i<j$.
Define $U(t_j)|\psi\rangle$ $=\,\sum\delta_r|\psi_r\rangle$,
$U(t_j)O_k(t_k)|\psi\rangle$ $=\,\sum\alpha^k_r|\psi_r\rangle$, and
$U(t_j)O_l(t_l)O_k(t_k)|\psi\rangle$ $=\, \sum\eta^{kl}_r|\psi_r\rangle$,
where $|\psi_r\rangle$ is any basis and $k,l\leq j$.
Expanding Eq. (\ref{expand1}) in the $|\psi_r\rangle$ basis up to the $j$th measurement yields
\begin{equation}
\label{simu}
\begin{split}
&|\Psi\rangle\!=\!\sum_r\Big[\delta_r\prod_k^j\!\phi(q_k)\!-\!\sum_k^j\!\alpha^k_r\phi'(q_k)\prod_{l\neq k}\! \phi(q_l)\!
\\&+\sum_{k<l}\!\phi'(q_k)\phi'(q_l)\!\prod_{s\neq k,l}\!\phi(q_s) \eta^{kl}_r
\\&+\frac{1}{2}\sum_k^j\!\phi''(q_k)\prod_{l\neq k} \!\phi(q_l)\eta^{kk}_r\!+\!o(\epsilon^3)\!\Big]\! |\psi_r\rangle.
\end{split}
\end{equation}
To compute $E_{QM}(q_iq_j)$ we trace out the system and $q_k$, $k\neq i\neq j$.
Using $\int q\phi^2(q)dq=0$, $\int\phi(q)\phi'(q)dq=0$,
it can be shown that to second order operations of $O_k$ do not contribute.
Therefore,
\begin{equation}
\label{seq0}
\begin{split}
E_{QM}(q_iq_j)&=\frac{1}{4}\sum_r(\delta_r^*\eta^{ij}_r+\delta_r\eta^{*ij}_r+\alpha^{*i}_r\alpha^j_r+\alpha^i_r\alpha^{*j}_r)\!+\!
o(\epsilon^4)
\\&=\mbox{Re}\langle\psi|O_i(t_i)O_j(t_j)|\psi\rangle+o(\epsilon^4).
\end{split}
\end{equation}
This completes the proof of theorem 1.
\\
\\
Using the same considerations it can be shown
that Eqs. (\ref{seq1},\ref{seq2}) are valid for simultaneous measurements too.
Related results for correlations of two-level system with continuous weak measurements have been discussed in \cite{korotkov},
the case of post-selection in \cite{steinberg,jozsa,stein2}
and two sequential measurements in \cite{johansen}.
\\
\\
{\bf Corollary 1.}
In the sequential setting we define
\begin{equation}
\label{chsh2}
B_S=|E(q^A_1 q^B_1)\!+\!E(q^A_1 q^B_2)\!+\!E(q^A_2 q^B_1)\!-\!E(q^A_2 q^B_2)|.
\end{equation}
In the limit of weak measurements $B_S\to B$,
coinciding with that defined
in Eq. (\ref{chsh}).
In particular, maximal violation of $2\sqrt{2}$ is obtained with the same observables as in the regular setting with
$O^A_1=\sigma^A_x\otimes I^B$,
$O^A_2=\sigma^A_z\otimes I^B$,
$O^B_1=I^A \otimes \sigma^B_{\pi/4}$ and
$O^B_2=I^A\otimes \sigma^B_{3\pi/4}$.
\\
\\
{\bf Corollary 2.} Assume that at the end of every cycle of weak measurements
the parties randomly choose a single observable to measure strongly $O^A_s,\, O^B_s$ respectively.
Then $E_{QM}(q^A_s,q^B_s)=\langle \psi|O^A_s \otimes O^B_s|\psi\rangle$.
This can be straightforwardly seen from the weak disturbance of the previously measured observables.
In quantum theory the weakly measured correlations equal the strongly measured ones.

\section{Testing local hidden variable models in the sequential setting}
In the sequential setting the outcomes of weak measurements may lie outside the usual range of the measured observables.
Then the hidden variables can interfere with the noise to violate CHSH inequality.
Therefore the sequential setting does not constitute an independent test of quantum nonlocality.
The requirement of random choice seems inevitable in any independent test of quantum nonlocality.

Nevertheless, physically one would expect additive random noise $n$: $q^A_i=a_i+n^A_i$, $q^B_j=b_j+n^B_j$
such that $E(a_i n_j)=0$ and $E(n_i n_j)_{i\neq j}=0$.
Then
\begin{equation}
\label{classic}
E(q^A_i q^B_j)=E(a_i b_j),
\end{equation}
and no violation of CHSH inequality can be shown.
By corollary 2 we can use the regular test of nonlocality
as a certificate for the sequential setting.
That is, let each party precisely measure a randomly chosen observable at the end of every cycle.
Bell's theorem states that the correlations of the precise measurements would not violate CHSH inequality.
If the parties observe that the weakly measured correlations differ from the precisely measured ones,
then they know that the sequential test is not valid.
However, then they also know that the hidden variables ("maliciously") interfere with the noise.
\section{Transition from strong to weak measurements}
\begin{figure}[ht]
\center{
\includegraphics[width=3.2in]{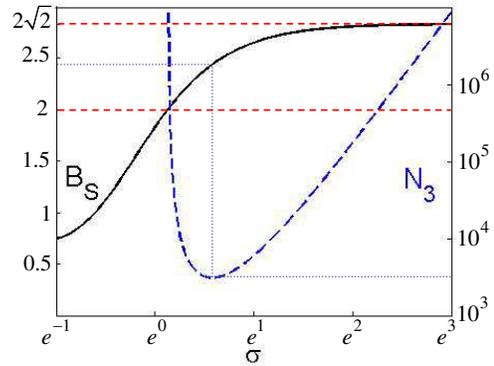}
\caption
{
Violation of the sequential CHSH test. Solid line: $B_S$
as a function of the measurements inaccuracy
$\sigma\equiv\Delta q(t=0)$.
$B_S>2$ for
$\sigma>\sqrt{[-2\ln(2^{3/4}-1)]^{-1}}\sim1.1425$.
Note that precise sequential measurements satisfy CHSH inequality.
Broken line: ensemble size $N_3(\sigma)$ that
manifests violation with three standard deviations.
$B_S\sim 2.43$ corresponds to the minimal ensemble size
$N_3=3088$. For comparison, the regular non-sequential precise setting requires
$N_3=105$.
}
\label{fig2}
}
\end{figure}

\begin{figure}[ht]
\center{
\includegraphics[width=3.2in]{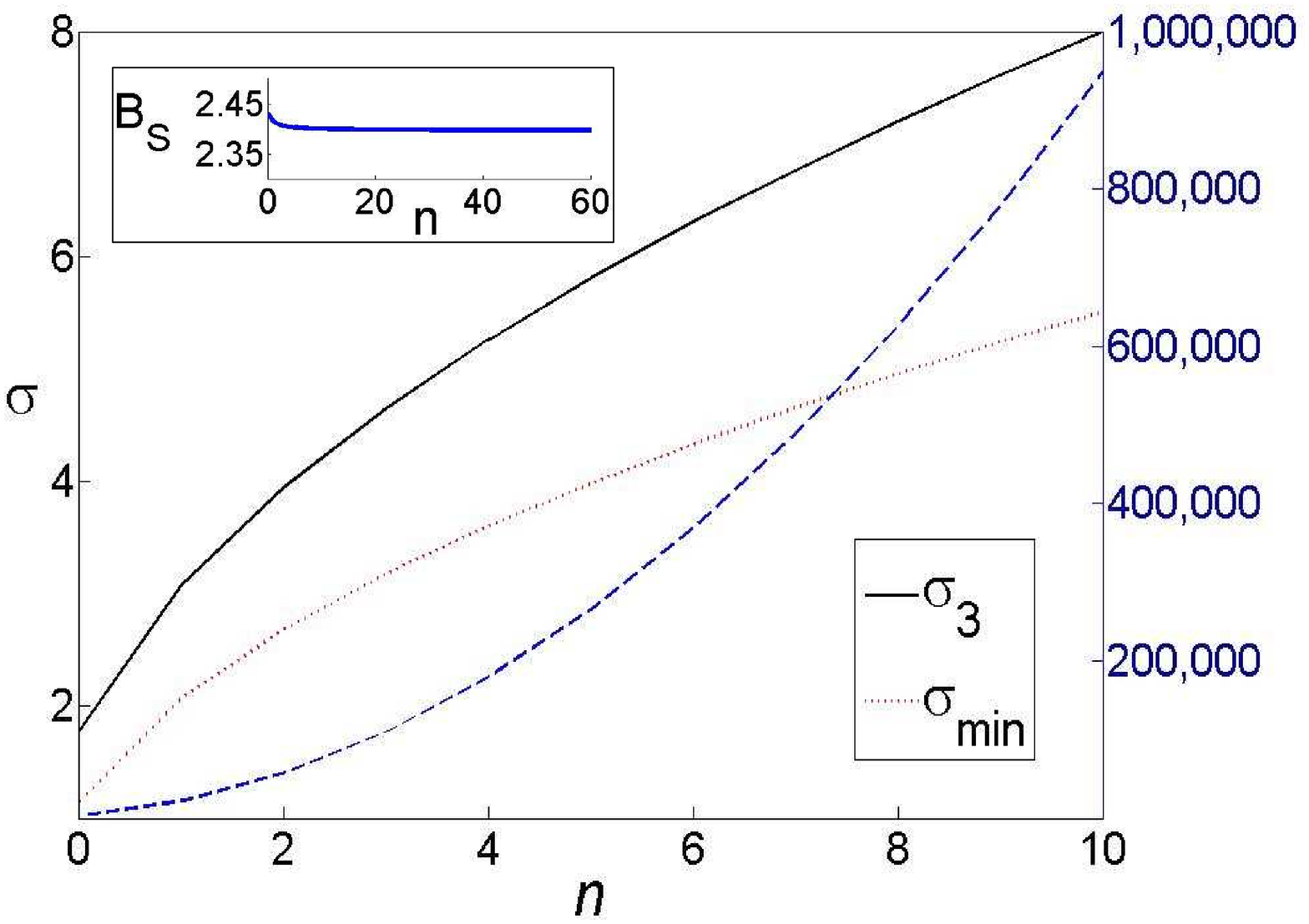}
\caption
{
Simultaneous violation of several CHSH tests.
The parties perform $n$ CHSH experiments before starting the current CHSH sequence.
$B_S>2$ for $\sigma>\sigma_{\mbox{min}}$ (dotted line),
where $\sigma_{\mbox{min}}=2^{3/4}\sqrt{n}$ (for large enough $n$).
Also shown is the minimal value of
$N_3$ (broken line) and the corresponding $\sigma_3$ (solid line)
which manifests violation with 3 standard deviations.
$N_3$ scales quadratically with $n$.
{\it Inset:} Expected violation given $\sigma_3$.
$B_S(n)$ decreases slowly with $n$ where $B_S\sim2.4$.
}
\label{fig3}
}
\end{figure}

The strict limit of weak measurements requires an infinite ensemble,
which is not practical.
However, violation can be observed using a finite ensemble, and
finite measurement apparatus inaccuracy
$\sigma\equiv\Delta q(t=0)$, 
as illustrated in fig. 2.
In fact, we find that $B_S>2$ for relatively mild value of $\sigma\sim1.15$.
As the inaccuracy increases $B_{S}\to 2\sqrt{2}$.
Also shown is the required ensemble size $N_3$ by which
violation is manifested with three standard deviations.
The minimal ensemble is about thirty times larger the one required to manifest $B>2$
in the usual setting.

Furthermore, simultaneous violation of (possibly different) Bell inequalities can be
observed using the same ensemble.
For example, assume that the two parties perform $n$ CHSH sequences
on each pair of the ensemble with a fixed inaccuracy $\sigma$ before starting the current CHSH test.
In fig. 3 we show the
minimal ensemble size $N_3$ and the corresponding inaccuracy
$\sigma_3$
by which violation is manifested.


We proceed by quantifying the required inaccuracy of the measurements as depicted in figures 2 and 3.
By explicitly expanding the joint state of the system and the measurement apparatuses
it can be shown that $E(q^A_1 q^B_1)=\frac{1}{\sqrt{2}}$,
$E(q^A_2 q^B_1)=E(q^A_1 q^B_2)=\frac{y}{\sqrt{2}}$
and $E(q^A_2 q^B_2)=\frac{y^2}{\sqrt{2}}$,
where
$y\equiv e^{-\epsilon/2}=e^{-\frac{1}{2\sigma^2}}$.
For example, the mutual state of the system and the measuring devices after party $A$
measures both
$\sigma^A_x$ and then $\sigma^A_z$ and party $B$ measures
$\sigma^B_{\pi/4}$ is
\begin{equation}
\begin{split}
&|\uparrow_z\uparrow_{\pi/4}\rangle\phi(q^A_2\!-\!1)\phi(q^B_1\!-\!1)\big(\alpha\phi(q^A_1\!-\!1)-\beta(q^A_1\!+\!1)\big)
\\+&|\downarrow_z\uparrow_{\pi/4}\rangle\phi(q^A_2\!+\!1)\phi(q^B_1\!-\!1)\big(\alpha\phi(q^A_1\!-\!1)+\beta(q^A_1\!+\!1)\big)
\\+&|\uparrow_z\downarrow_{\pi/4}\rangle\phi(q^A_2\!-\!1)\phi(q^B_1\!+\!1)\big(\beta\phi(q^A_1\!-\!1)+\alpha(q^A_1\!+\!1)\big)
\\+&|\downarrow_z\downarrow_{\pi/4}\rangle\phi(q^A_2\!+\!1)\phi(q^B_1\!+\!1)\big(\beta\phi(q^A_1\!-\!1)-\alpha(q^A_1\!+\!1)\big),
\end{split}
\nonumber
\end{equation}
where
$\alpha=\frac{1}{2\sqrt{2}}[\cos(\pi/8)+\sin(\pi/8)]$,
$\beta=\frac{1}{2\sqrt{2}}[\sin(\pi/8)-\cos(\pi/8)]$.
By tracing out the system and $q_1^A$, one can compute $E(q_2^A q_1^B)$ given above.
Therefore $B_{S}=\frac{1}{\sqrt{2}}(1+y)^2$, where $B_S$ denotes the sequential setting.

By the central limit theorem the size of the ensemble which is required to manifest
violation of Bell inequality with $z$ standard deviations is determined by
\begin{equation}
z\sqrt{\frac{V(B)}{N_3}}<B-2,
\end{equation}
where $V(B)$ is the variance of $B$.
In the regular non-sequential setting $B$ is the sum of four independent variables
$V(B)\equiv\sum_{i,j} V(q^A_iq^B_j)=4(1+\sigma^2)^2-2$.
In case of sequential measurements the experiments are dependent,
thus
$V(B_S)\equiv\sum_{i,j,k,l}CV(q^A_i q^B_j,q^A_k q^B_l)$,
where
$CV(x,y)=E(xy)-E(x)E(y)$ is the mutual covariance.
By a straightforward computation it can be shown that
for $i\neq k$ and $j\neq l$,
$CV(q^A_i a^B_j,q^A_i a^B_l)=CV(q^A_i q^B_j,q^A_k q^B_l)=0$,
and $V(q^A_i q^B_j)=(1+\sigma^2)^2$.
Therefore
$V(B_S)=4(1+\sigma^2)^2-\frac{1}{2}(1+y)^4$.
These results are illustrated in fig. 2.

Now assume that the two parties perform $n$ CHSH sequences before starting the current CHSH
experiment, where the $i$th observable of each party is maximally non-commuting with the $i-1$ one for any $i$.
Then it can be shown that
\begin{equation}
\label{subseq1}
B_{S}(n)=\frac{y^{2n}}{\sqrt{2}}(1+y)^2
\end{equation}
and $V\big(B_S(n)\big)=4(1+\sigma^2)^2-\frac{y^{2n}}{2}(1+y)^4$.
These results are illustrated in fig. 3.
It can be numerically shown that Eq. (\ref{subseq1})
yields a typical value for randomly chosen $n=2m$ measurements.

\section{Maximal violation of Leggett-Garg inequalities}
Our suggestion to manifest
violation of Bell inequalities is related to the time inequalities first
suggested by Leggett and Garg (LG) \cite{leggett}.
LG discuss measurements on a local system
which are performed at different times.
They analyze the scenario in which at each cycle the experimentalist chooses
two out of $m>2$ measurements and compute their mutual expectation value.
Given a realistic model and non-demolition measurements,
correlations satisfy the LG inequalities, which have a similar structure to Bell inequalities.
Interestingly, it was shown that quantum theory violates the LG inequalities
using weak measurements \cite{mizel,jordan1,jordan} even if
all observables are measured on each member of the ensemble.
In this setting precise measurements
do not violate LG inequalities by definition.
The addition of noise, however, distinguishes models incorporating LG assumptions from quantum theory,
since non-demolition measurements imply that the noise and the measured observable are independent \cite{mizel}.
This is violated in quantum theory even though the effect of weak measurements on the system is small.
An immediate consequence of Eq. (\ref{seq2}) is that LG inequalities are maximally violated by quantum theory
in the limit of weak measurements.
Note that the realism assumption is crucial since one can write a joint probability distribution to
the pointers of weak measurements,
which do not correspond to $\pm1$ values.

\section{Conclusions} We propose a new approach to test quantum nonlocality,
in which all observables are measured on each system without the need
to select randomly a single observable.
By enlarging the sequence of measurements it is also possible to manifest a simultaneous
violation of more than one Bell inequality.
Our proposal can be demonstrated in all systems
in which both Bell state preparation and sequential weak measurements are realizable,
for example atomic physics devices \cite{ion}, photons \cite{amplify,photon}, and
solid-state qubits \cite{NP}.
%
\\
\\
\acknowledgments
We thank P. Skrzypczyk, N. Klinghoffer, J. Kupferman and G. Ben-Porath for helpful discussions.
This work has been supported by the Israel Science Foundation grant numbers 784/06, 920/09.

\end{document}